\documentclass[12pt]{article}
\usepackage[top = 2 cm, bottom = 2.5 cm, left = 2 cm, right = 2 cm]{geometry}
\usepackage{authblk, cite, color, amssymb, amsmath}
\usepackage[toc]{appendix}
\usepackage[hypertex, colorlinks = true, linkcolor = blue, citecolor = red]{hyperref}
\usepackage[dvipsnames]{xcolor}

\usepackage{mathrsfs}
\usepackage{amsmath}
\usepackage{amssymb}

\begin{document}

\title{Dirac and Maxwell systems in split octonions}

\author{\bf Merab Gogberashvili}

\affil{{\small Javakhishvili Tbilisi State University, 3 Chavchavadze Avenue, Tbilisi 0179, Georgia \authorcr
Andronikashvili Institute of Physics, 6 Tamarashvili Street, Tbilisi 0177, Georgia} \authorcr
e-mail: gogber@gmail.com}

\author{\bf Alexandre Gurchumelia}

\affil{{\small Javakhishvili Tbilisi State University, 3 Chavchavadze Avenue, Tbilisi 0179, Georgia \authorcr
Kharadze Georgian National Astrophysical Observatory, Abastumani 0301, Georgia} \authorcr
e-mail: alexandre.gurchumelia@gmail.com}

\maketitle

\begin{abstract}

The known equivalence of 8-dimensional chiral spinors and vectors, also referred to as triality, is discussed for (4+4)-space. Split octonionic representation of SO(4,4) and Spin(4,4) groups and the trilinear invariant form are explicitly written and compared with Clifford algebraic matrix representation. It is noted that the complete algebra of split octonionic basis units can be recovered from the Moufang and Malcev relations for the three vector-like elements. Lagrangians on split octonionic fields that generalize Dirac and Maxwell systems are constructed using group invariant forms. It is shown that corresponding equations are related to split octonionic analyticity conditions.

\vskip 2mm
\noindent
Subject class: Primary 17A35,
15A66;
Secondary 34L40,
35Q61
\vskip 1mm
\noindent
Keywords: Split octonions, Triality, SO(4,4), Dirac and Maxwell equations
\end{abstract}


\tableofcontents


\section{Introduction}

Nonassociative algebras, apart from Lie algebras, have never been
systematically utilized in physics, only some attempts have been made.
Nevertheless, there are some intriguing hints that these kinds of
algebras may play an essential role in the ultimate theory, yet to
be discovered. Octonions, as an example of such a nonassociative structure,
form the largest normed division algebra after the algebras of real,
complex and quaternionic numbers \cite{Sc,Sp-Ve,Baez}. Since their
discovery in 1844-1845 there have been various attempts to find appropriate
uses for octonions in physics (see reviews \cite{Rev-1,Rev-3,Rev-4,Rev-2}).
One can point to the possible impact of octonions on: Color symmetry
\cite{Color-1,Color-2}; GUTs \cite{GUT-1,GUT-2,GUT-3}; Representation
of Clifford algebras \cite{Cliff-1,Cliff-2}; Quantum mechanics \cite{QM-1,QM-2};
Space-time symmetries \cite{Rel-1,Rel-2,Gogberashvili:2022kzw}; Formulations
of wave equations \cite{WE-1,WE-2,WE-3}; Quantum Hall effect \cite{Hall};
Kaluza-Klein program without extra dimensions \cite{KK-1,KK-2}; Strings
and M-theory \cite{String-1,String-2,String-2',String-3,String-4};
SUSY \cite{SUSY-1,SUSY-2,Schray:1994up,Anastasiou:2013cya}; etc.

Eight-dimensional Euclidean space, in which ordinary octonions reside,
possesses certain peculiarities, namely that the vector and the two
chiral parts of the spinor are all eight-dimensional objects and there
exists a rotation invariant trilinear form in which vectors and chiral
spinors act indistinguishably from one another. This property, called
triality \cite{Gamba,Dray}, is usually formulated in terms of spin
group automorphisms and symmetry of $D_{4}$ Dynkin diagram \cite{Cliff-1}.

Properties of spinors and vectors have been also discussed within
the context of split octonions. Unlike ordinary octonions, the split
algebra lacks the property of divisibility since it contains zero
divisors. On the other hand (4+4)-space of the split octonions has
Minkowskian subspaces. Hence $SO(8)$ group describing rotational
symmetry of the Euclidean space is replaced by its noncompact analog
for (4+4)-space, namely $SO(4,4)$, in which Lorentz groups $SO(1,3)$
and $SO(3,1)$ are contained multiple times as subgroups. This makes
the split octonions interesting to study in the context of geometry
in physics \cite{Gog-1,Gog-2,Go-Sa,Go-Gu}.

In physical applications split octonions were used to provide possible
explanation for the existence of three generations of fermionic elementary
particles \cite{Gunaydin:1974fb,Silagadze}. In \cite{PBN} generators
of $SO(8)$ and $SO(7)$ groups were obtained and have been used to
describe the rotational transformation in 7-dimensional space. In
\cite{Nash-1,Nash-2} real reducible $16\times16$ matrix representation
of $SO(4,4)$ group utilizing the Clifford algebra approach was constructed
and it was shown that there are two inequivalent real $8\times8$
irreducible basic spinor representations, potential implementation
for 8-dimensional electrodynamics \cite{Nash-2} and gravity \cite{Nash-3}
was also considered. In \cite{DeAndrade:2000jx} the basic features
of Cartan's triality of $SO(8)$ and $SO(4,4)$ was analyzed in the
Majorana-Weyl basis, it was shown that the three Majorana-Weyl spacetimes
of signatures (4+4), (8+0), (0+8) are interrelated via the permutation
group (signature-triality). In \cite{Dray} octonionic representation
of $SO(8)$ and triality was discussed, but triality symmetry is also
valid in (4+4)-space spanned by the split octonion algebra. Another
unique concept associated only with (4+4)-space is 4-ality, it's similar
to triality but deals with fourfold symmetry of modified Dynkin diagram
$\tilde{D}_{4}$ \cite{Landsberg}.

One of the objectives of this article is to recast results of \cite{Gamba}
to (4+4)-space. We also want to describe split octonionic vectorial
and spinorial representations of $SO(4,4)$ group and to construct
split octonionic Dirac and Maxwell Lagrangians underlying triality
symmetry in this space.

The paper is organized as follows. In Sec.~\ref{sec:Cl(4,4)} we
present $16\times16$ complex matrix representation of the Clifford
algebra $\mathcal{C}\ell_{4,4}$. The Sec.~\ref{sec:vectors} and
Sec.~\ref{sec:spinors} are devoted to vectorial and spinorial matrix
representations of $SO(4,4)$ group respectively. In the Sec.~\ref{sec:triality}
the equivalence of (4+4)-vectors and chiral spinors (triality) is
explicitly demonstrated. In the Sec.~\ref{split} it is shown that
the complete algebra of hypercomplex octonionic basis units can be
recovered from the Moufang and Malcev relations. In Sec.~\ref{sec:Trilinear}
the trilinear form, and the group $SO(4,4)$ under which it is invariant,
is written in terms of split octonions. In Sec.~\ref{sec:Field_theory}
split octonionic Lagrangians that can be built by quadratic and trilinear
invariant forms are presented. In the Sec.~\ref{sec:Dirac+Maxwell}
it is shown that an equation similar to the split octonionic analyticity
condition can be reduced to the system of the Dirac-Maxwell equations.
Finally, Sec.~\ref{sec:conclusions} presents our conclusions.


\section{Matrix representation of $\mathcal{C}\ell_{4,4}$}

\label{sec:Cl(4,4)} \setcounter{equation}{0}

Geometric algebra of (4+4)-space is a Clifford algebra over the real
number field with a diagonal metric $g_{\mu\nu}$ (Greek indices,
e.g. $\mu,\nu$ take on the values $0,1,\ldots,7$) having $(4,4)$
signature and is usually denoted as $\mathcal{C}\ell_{4,4}$. As all
Clifford algebras, $\mathcal{C}\ell_{4,4}$ is associative and can
be defined through anti-commutation relations: 
\begin{equation}
e_{\mu}e_{\nu}+e_{\nu}e_{\mu}=2g_{\mu\nu}~,\label{eq:Cl44_algebra}
\end{equation}
where $e_{\mu}$ are orthogonal basis units of grade-1 vectors.

Basis unit $e_{\mu}$ can be represented as $\Gamma_{\mu}$-matrix.
To obtain an exact form of the $\Gamma_{\mu}$-matrices for $\mathcal{C}\ell_{4,4}$,
we can take the $\mathcal{C}\ell_{8,0}$ generating matrices $A_{\mu}$
described in \cite{Gamba} and multiply four of them by complex imaginary
unit $i$, 
\begin{equation}
\begin{split}\Gamma_{\mu} & =A_{\mu}~,\qquad\left(\mu=0,1,2,3\right)\\
\Gamma_{\nu} & =iA_{\nu}~.\qquad\left(\nu=4,5,6,7\right)
\end{split}
\label{Gamma}
\end{equation}
This changes the Euclidean metric into the split metric of (4+4)-space.
Here we use labeling and ordering of 16-dimensional Hermitian $A_{\mu}$
matrices that differs from the one in \cite{Gamba}, 
\begin{equation}
A_{\mu}=\left(\begin{array}{cc}
0 & \alpha_{\mu}\\
\alpha_{\mu}^{\dagger} & 0
\end{array}\right)~,
\end{equation}
where the 8-dimensional $\alpha_{\mu}$ matrices are: 
\[
\alpha_{0}=\left(\begin{array}{cc|cc|cc|cc}
-1 &  &  &  &  & \\
 & 1 &  &  &  & \\
\hline  &  & 1 &  &  & \\
 &  &  & 1 &  & \\
\hline  &  &  &  & -1 & \\
 &  &  &  &  & -1\\
\hline  &  &  &  &  &  & -1\\
 &  &  &  &  &  &  & 1
\end{array}\right),\quad\alpha_{1}=\left(\begin{array}{cc|cc|cc|cc}
i &  &  &  &  & \\
 & i &  &  &  & \\
\hline  &  & i &  &  & \\
 &  &  & i &  & \\
\hline  &  &  &  & i & \\
 &  &  &  &  & i\\
\hline  &  &  &  &  &  & i\\
 &  &  &  &  &  &  & i
\end{array}\right),
\]

\[
\alpha_{2}=\left(\begin{array}{cc|cc|cc|cc}
 & 1 &  &  &  & \\
1 &  &  &  &  & \\
\hline  &  &  &  & -1 & \\
 &  &  &  &  & -1\\
\hline  &  & -1 &  &  & \\
 &  &  & -1 &  & \\
\hline  &  &  &  &  &  &  & 1\\
 &  &  &  &  &  & 1
\end{array}\right),\quad\alpha_{3}=\left(\begin{array}{cc|cc|cc|cc}
 & -i &  &  &  & \\
i &  &  &  &  & \\
\hline  &  &  &  & i & \\
 &  &  &  &  & i\\
\hline  &  & -i &  &  & \\
 &  &  & -i &  & \\
\hline  &  &  &  &  &  &  & -i\\
 &  &  &  &  &  & i
\end{array}\right),
\]

\[
\alpha_{4}=\left(\begin{array}{cc|cc|cc|cc}
 &  & 1 &  &  & \\
 &  &  &  & 1 & \\
\hline 1 &  &  &  &  & \\
 &  &  &  &  &  & -1\\
\hline  & 1 &  &  &  & \\
 &  &  &  &  &  &  & -1\\
\hline  &  &  & -1 &  & \\
 &  &  &  &  & -1
\end{array}\right),\quad\alpha_{5}=\left(\begin{array}{cc|cc|cc|cc}
 &  & i &  &  & \\
 &  &  &  & i & \\
\hline -i &  &  &  &  & \\
 &  &  &  &  &  & -i\\
\hline  & -i &  &  &  & \\
 &  &  &  &  &  &  & -i\\
\hline  &  &  & i &  & \\
 &  &  &  &  & i
\end{array}\right),
\]

\[
\alpha_{6}=\left(\begin{array}{cc|cc|cc|cc}
 &  &  & 1 &  & \\
 &  &  &  &  & 1\\
\hline  &  &  &  &  &  & 1\\
1 &  &  &  &  & \\
\hline  &  &  &  &  &  &  & 1\\
 & 1 &  &  &  & \\
\hline  &  & 1 &  &  & \\
 &  &  &  & 1 & 
\end{array}\right),\quad\alpha_{7}=\left(\begin{array}{cc|cc|cc|cc}
 &  &  & -i &  & \\
 &  &  &  &  & -i\\
\hline  &  &  &  &  &  & -i\\
i &  &  &  &  & \\
\hline  &  &  &  &  &  &  & -i\\
 & i &  &  &  & \\
\hline  &  & i &  &  & \\
 &  &  &  & i & 
\end{array}\right).
\]

We see that four of the $\alpha_{\mu}$ matrices, and thus four corresponding
$A_{\mu}$ matrices, are imaginary and the rest four are real. For
obtaining (4+4)-space algebra from that Euclidean version, we could
have chosen any four of the eight generators to be multiplied by the
complex imaginary unit $i$. Choosing the imaginary $A_{\mu}$ matrices
would have resulted in a real representation of $\mathcal{C}\ell_{4,4}$,
which is indeed \textit{algebra isomorphic} to the ring of $16\times16$
real matrices \cite{Cliff-1}. However, in the complex representation
defined above (\ref{Gamma}), some calculations are easier and closer
to those provided for Euclidean 8-space in \cite{Gamba}.


\section{Vectors in (4+4)-space}

\label{sec:vectors} \setcounter{equation}{0}

Let us take $x$ to be a real vector in (4+4)-space whose components
are labeled as $x_{\mu}$. Object that \textit{transforms like a vector}
is represented by a matrix 
\begin{equation}
x=\sum_{\beta=0}^{7}x_{\beta}\Gamma_{\beta}~,\label{eq:cl44vector}
\end{equation}
where $\Gamma_{\beta}$-matrices are defined in (\ref{Gamma}). The
vectors of (4+4)-space have the property that 
\begin{equation}
x^{2}=x_{0}^{2}+x_{1}^{2}+x_{2}^{2}+x_{3}^{2}-x_{4}^{2}-x_{5}^{2}-x_{6}^{2}-x_{7}^{2}~,\label{eq:cl44quadratic_form}
\end{equation}
where we assume that the right-hand side is multiplied by the $(16\times16)$
identity matrix.

The similarity transformations 
\begin{equation}
x^{\prime}=L_{\mu\nu}\left(\vartheta\right)xL_{\mu\nu}^{-1}\left(\vartheta\right)~,\label{eq:vec_transform}
\end{equation}
where 
\begin{equation}
L_{\mu\nu}\left(\vartheta\right)=\exp\left(-\frac{1}{2}\vartheta\Gamma_{\mu}\Gamma_{\nu}\right)~,\label{L}
\end{equation}
result in rotations and boosts of the vector $x$. This represents
the $SO(4,4)$ group under which the quadratic form (\ref{eq:cl44quadratic_form})
is invariant. Transformations of $x$ under $L_{\mu\nu}$ can be divided
into two types: one comprising 2 copies of $SO(4)$ in two maximal
anisotropic subspaces and another comprising 16 copies of $SO(1,1)$
that mix these two maximal anisotropic subspaces in isotropic planes.
The former type of transformations is compact and they are realized
when either $\mu,\nu=0,1,2,3$ or $\mu,\nu=4,5,6,7$. The latters
are Lorentz-like non-compact boosts, i.e. hyperbolic transformations
and are realized when $\mu=0,1,2,3$ and $\nu=4,5,6,7$, or vice versa.

To demonstrate these two different types of $SO(4,4)$-transformations,
it is sufficient to study them in the tangential space. The space
is spanned by Taylor expansion of the transformation matrix (\ref{L})
in the neighborhood of the identity element up to the first order
term, 
\begin{equation}
L_{\mu\nu}\left(\vartheta\right)\simeq1-\frac{1}{2}\vartheta\Gamma_{\mu}\Gamma_{\nu}~.
\end{equation}
Using the fact that 
\begin{equation}
L_{\mu\nu}^{-1}=L_{\nu\mu}~,
\end{equation}
the formula (\ref{eq:vec_transform}) in the tangential space reduces
to 
\begin{equation}
x^{\prime}=\sum_{\alpha}x_{\alpha}^{\prime}\Gamma_{\alpha}=\sum_{\beta}\left[x_{\beta}\Gamma_{\beta}-\frac{1}{2}\vartheta x_{\beta}\left(\Gamma_{\mu}\Gamma_{\nu}\Gamma_{\beta}+\Gamma_{\beta}\Gamma_{\nu}\Gamma_{\mu}\right)\right]~.\label{eq:vec_transform_detailed}
\end{equation}

As an example let us consider rotations in $e_{4}\wedge e_{5}$ plane.
For $\beta\neq4,5$ the second term in (\ref{eq:vec_transform_detailed})
vanishes due to the defining algebraic relation (\ref{eq:Cl44_algebra}),
so we can write 
\begin{equation}
x_{\beta}^{\prime}=x_{\beta}~.
\end{equation}
When $\beta=5$ the second term in (\ref{eq:vec_transform_detailed})
turns into $\vartheta x_{5}\Gamma_{4}$, which dictates that 
\begin{equation}
x_{4}^{\prime}=x_{4}+\vartheta x_{5}~.
\end{equation}
Similarly, for $\beta=4$ we get 
\begin{equation}
x_{5}^{\prime}=x_{5}-\vartheta x_{4}~.
\end{equation}
Since we have opposite sign in front of $\vartheta$ in these two
infinitesimal coordinate transformations, corresponding finite transformations
would result in compact rotations: 
\begin{equation}
\begin{split}x_{4}^{\prime} & =x_{4}\cos\vartheta+x_{5}\sin\vartheta~,\\
x_{5}^{\prime} & =x_{5}\cos\vartheta-x_{4}\sin\vartheta~,\\
x_{\rho}^{\prime} & =x_{\rho}~.\qquad\qquad(\rho\neq4,5)
\end{split}
\end{equation}
We have similar compact rotations in all anisotropic planes.

Alternatively, the transformations that mix maximal anisotropic subspaces
are non-compact. For example, if we apply calculations similar to
the previous case to $\mu=0$ and $\nu=4$, we would get non-compact
rotations of the form: 
\begin{equation}
\begin{split}x_{0}^{\prime} & =x_{0}\cosh\vartheta+x_{4}\sinh\vartheta~,\\
x_{4}^{\prime} & =x_{4}\cosh\vartheta+x_{0}\sinh\vartheta~,\\
x_{\rho}^{\prime} & =x_{\rho}~.\qquad\qquad(\rho\neq0,4)
\end{split}
\end{equation}

At the end of this section we want to introduce one of the $1680$
possible grade-$4$ elements of $\mathcal{C}\ell_{4,4}$, 
\begin{equation}
B=-\Gamma_{1}\Gamma_{3}\Gamma_{5}\Gamma_{7}~,
\end{equation}
which due to the property 
\begin{equation}
\Gamma_{\mu}^{T}=B\Gamma_{\mu}B~,\qquad\left(\mu=0,1,\ldots,7\right)\label{eq:B_property}
\end{equation}
will become useful below.


\section{Spinors in (4+4)-space}

\label{sec:spinors} \setcounter{equation}{0}

A spinor in the (4+4)-space can be represented as a $16$-dimensional
column vector 
\begin{equation}
\eta=\phi+\psi~,\label{eq:eta}
\end{equation}
where 
\begin{equation}
\begin{array}{cc}
\phi=\left(\begin{array}{c}
\phi_{0}\\
\phi_{1}\\
\vdots\\
\phi_{7}\\
0\\
0\\
\vdots\\
0
\end{array}\right)\quad\text{and}\thickspace & \psi=\left(\begin{array}{c}
0\\
0\\
\vdots\\
0\\
\psi_{0}\\
\psi_{1}\\
\vdots\\
\psi_{7}
\end{array}\right)\end{array}\label{eq:cl44chiral_spinors}
\end{equation}
are spinors of different chirality. We note that $\phi$ and $\psi$
have 8 independent real components each.

The spinor transformations under $Spin(4,4)$ (double cover of $SO(4,4)$)
are described by the same matrix (\ref{L}) that was used for vectors,
but the transformation law is different 
\begin{equation}
\eta^{\prime}=L_{\mu\nu}\left(\vartheta\right)\eta~.\label{eq:spin_transform}
\end{equation}
Under this transformation the quantity 
\begin{equation}
\eta^{T}B\eta=\phi^{T}B\phi+\psi^{T}B\psi\label{eq:spinor_invariant}
\end{equation}
is invariant. We prove this in the tangential space using the property
of $B$ matrix (\ref{eq:B_property}), 
\begin{equation}
\begin{split}\eta^{\prime}{}^{T}B\eta^{\prime} & =\eta^{T}\left(1+\frac{1}{2}\vartheta\Gamma_{\nu}^{T}\Gamma_{\mu}^{T}\right)B\left(1+\frac{1}{2}\vartheta\Gamma_{\mu}\Gamma_{\nu}\right)\eta=\\
 & =\eta^{T}B\left(1-\frac{1}{2}\vartheta\Gamma_{\mu}\Gamma_{\mu}\right)\left(1+\frac{1}{2}\vartheta\Gamma_{\mu}\Gamma_{\nu}\right)\eta=\eta^{T}B\eta~.
\end{split}
\end{equation}
It can be noticed that two terms on the right hand side of the relation
(\ref{eq:spinor_invariant}) are conserved independently, meaning
that their terms do not mix.


\section{Triality}

\label{sec:triality} \setcounter{equation}{0}

The vector $x$ considered in the Sec.~\ref{sec:vectors} and two
kind of spinors $\psi$ and $\phi$ considered in the Sec.~\ref{sec:spinors}
are objects of same dimension in the underlying field $\mathbb{R}$.
This kind of match between the dimensions of vector and chiral spinors
only takes place in 8-dimensional space.

In order to extract another peculiarity of (4+4)-space, which relies
on the previous one, let us apply the following linear basis change
to the spinor (\ref{eq:eta}): 
\begin{equation}
\xi=\frac{1}{\sqrt{2}}\left(\begin{array}{c}
-\phi_{2}+i\phi_{3}\\
\thickspace\phi_{0}-i\phi_{1}\\
-\phi_{7}-i\phi_{6}\\
-\phi_{5}+i\phi_{4}\\
-\phi_{5}-i\phi_{4}\\
\thickspace\phi_{7}-i\phi_{6}\\
-\phi_{0}-i\phi_{1}\\
-\phi_{2}-i\phi_{3}\\
\thickspace\psi_{2}-i\psi_{3}\\
-\psi_{0}-i\psi_{1}\\
-\psi_{7}-i\psi_{6}\\
-\psi_{5}+i\psi_{4}\\
\thickspace\psi_{5}+i\psi_{4}\\
-\psi_{7}+i\psi_{6}\\
-\psi_{0}+i\psi_{1}\\
-\psi_{2}-i\psi_{3}
\end{array}\right)~.
\end{equation}
In this basis, the invariant quadratic form (\ref{eq:spinor_invariant})
for 8-spinors $\phi$ and $\psi$ yields 
\begin{equation}
\begin{aligned}\phi^{T}B\phi=\thinspace & \phi_{0}^{2}+\phi_{1}^{2}+\phi_{2}^{2}+\phi_{3}^{2}-\phi_{4}^{2}-\phi_{5}^{2}-\phi_{6}^{2}-\phi_{7}^{2}~,\\
\psi^{T}B\psi=\thinspace & \psi_{0}^{2}+\psi_{1}^{2}+\psi_{2}^{2}+\psi_{3}^{2}-\psi_{4}^{2}-\psi_{5}^{2}-\psi_{6}^{2}-\psi_{7}^{2}~,
\end{aligned}
\label{eq:cl44spinors_quadratic_forms}
\end{equation}
which are analogous to the invariant quadratic form for the vector
(\ref{eq:cl44quadratic_form}). Then one can construct a trilinear
form 
\begin{equation}
\mathcal{F}:\mathbb{R}^{8}\times\mathbb{R}^{8}\times\mathbb{R}^{8}\rightarrow\mathbb{R}
\end{equation}
on $x$, $\phi$ and $\psi$, 
\begin{equation}
\mathcal{F}\left(\phi,x,\psi\right)=\phi^{T}Bx\psi~,\label{F_matrixrep}
\end{equation}
which is preserved under simultaneously transforming $x$ and $\eta=\phi+\psi$
under the vector (\ref{eq:vec_transform}) and spinor (\ref{eq:spin_transform})
transformation rules with the same $L_{\mu\nu}$. Proof is provided
in the tangential space: 
\begin{equation}
\begin{split}\phi^{\prime}{}^{T}Bx^{\prime}\psi^{\prime} & =\phi{}^{T}L_{\mu\nu}^{T}BL_{\mu\nu}xL_{\nu\mu}L_{\mu\nu}\psi=\\
 & =\phi{}^{T}\left(1+\frac{1}{2}\vartheta\Gamma_{\nu}^{T}\Gamma_{\mu}^{T}\right)B\left(1+\frac{1}{2}\vartheta\Gamma_{\mu}\Gamma_{\nu}\right)x\psi=\phi{}^{T}Bx\psi~.
\end{split}
\end{equation}

Let us look closely at these transformations. For example, the infinitesimal
$L_{01}\left(\vartheta\right)$ rotations of vector and spinors are:
\begin{equation}
\left\{ \begin{aligned}x_{0}^{\prime}= & \thinspace x_{0}-\vartheta x_{1}\\
x_{1}^{\prime}= & \thinspace x_{1}+\vartheta x_{0}\\
x_{2}^{\prime}= & \thinspace x_{2}\\
x_{3}^{\prime}= & \thinspace x_{3}\\
x_{4}^{\prime}= & \thinspace x_{4}\\
x_{5}^{\prime}= & \thinspace x_{5}\\
x_{6}^{\prime}= & \thinspace x_{6}\\
x_{7}^{\prime}= & \thinspace x_{7}
\end{aligned}
\right.~,\qquad\left\{ \begin{aligned}\phi_{0}^{\prime}= & \thinspace\phi_{0}+\tfrac{1}{2}\vartheta\phi_{1}\\
\phi_{1}^{\prime}= & \thinspace\phi_{1}-\tfrac{1}{2}\vartheta\phi_{0}\\
\phi_{2}^{\prime}= & \thinspace\phi_{2}-\tfrac{1}{2}\vartheta\phi_{3}\\
\phi_{3}^{\prime}= & \thinspace\phi_{3}+\tfrac{1}{2}\vartheta\phi_{2}\\
\phi_{4}^{\prime}= & \thinspace\phi_{4}-\tfrac{1}{2}\vartheta\phi_{5}\\
\phi_{5}^{\prime}= & \thinspace\phi_{5}+\tfrac{1}{2}\vartheta\phi_{4}\\
\phi_{6}^{\prime}= & \thinspace\phi_{6}+\tfrac{1}{2}\vartheta\phi_{7}\\
\phi_{7}^{\prime}= & \thinspace\phi_{7}-\tfrac{1}{2}\vartheta\phi_{6}
\end{aligned}
\right.~,\qquad\left\{ \begin{aligned}\psi_{0}^{\prime}= & \thinspace\psi_{0}+\tfrac{1}{2}\vartheta\psi_{1}\\
\psi_{1}^{\prime}= & \thinspace\psi_{1}-\tfrac{1}{2}\vartheta\psi_{0}\\
\psi_{2}^{\prime}= & \thinspace\psi_{2}+\tfrac{1}{2}\vartheta\psi_{3}\\
\psi_{3}^{\prime}= & \thinspace\psi_{3}-\tfrac{1}{2}\vartheta\psi_{2}\\
\psi_{4}^{\prime}= & \thinspace\psi_{4}+\tfrac{1}{2}\vartheta\psi_{5}\\
\psi_{5}^{\prime}= & \thinspace\psi_{5}-\tfrac{1}{2}\vartheta\psi_{4}\\
\psi_{6}^{\prime}= & \thinspace\psi_{6}-\tfrac{1}{2}\vartheta\psi_{7}\\
\psi_{7}^{\prime}= & \thinspace\psi_{7}+\tfrac{1}{2}\vartheta\psi_{6}
\end{aligned}
\right.~.\label{eq:L_01}
\end{equation}
As usual one full rotation for a vector $x$ is only half a rotation
for spinors $\phi$ and $\psi$. However, since $L_{\mu\nu}$ matrices
form a group under matrix multiplication, we can construct transformations
for $x$ that exactly reproduce transformations (\ref{eq:L_01}) of
$\phi$, 
\begin{equation}
\begin{aligned} & L_{10}\left(\frac{\vartheta}{2}\right)L_{23}\left(\frac{\vartheta}{2}\right)L_{54}\left(\frac{\vartheta}{2}\right)L_{67}\left(\frac{\vartheta}{2}\right)\simeq\\
 & \simeq1-\frac{1}{4}\vartheta\left(\Gamma_{1}\Gamma_{0}+\Gamma_{2}\Gamma_{3}+\Gamma_{5}\Gamma_{4}+\Gamma_{6}\Gamma_{7}\right)~,
\end{aligned}
\end{equation}
which results in 
\begin{equation}
\left\{ \begin{aligned}x_{0}^{\prime}= & \thinspace x_{0}+\tfrac{1}{2}\vartheta x_{1}\\
x_{1}^{\prime}= & \thinspace x_{1}-\tfrac{1}{2}\vartheta x_{0}\\
x_{2}^{\prime}= & \thinspace x_{2}-\tfrac{1}{2}\vartheta x_{3}\\
x_{3}^{\prime}= & \thinspace x_{3}+\tfrac{1}{2}\vartheta x_{2}\\
x_{4}^{\prime}= & \thinspace x_{4}-\tfrac{1}{2}\vartheta x_{5}\\
x_{5}^{\prime}= & \thinspace x_{5}+\tfrac{1}{2}\vartheta x_{4}\\
x_{6}^{\prime}= & \thinspace x_{6}+\tfrac{1}{2}\vartheta x_{7}\\
x_{7}^{\prime}= & \thinspace x_{7}-\tfrac{1}{2}\vartheta x_{6}
\end{aligned}
\right.~,\qquad\left\{ \begin{aligned}\phi_{0}^{\prime}= & \thinspace\phi_{0}+\tfrac{1}{2}\vartheta\phi_{1}\\
\phi_{1}^{\prime}= & \thinspace\phi_{1}-\tfrac{1}{2}\vartheta\phi_{0}\\
\phi_{2}^{\prime}= & \thinspace\phi_{2}+\tfrac{1}{2}\vartheta\phi_{3}\\
\phi_{3}^{\prime}= & \thinspace\phi_{3}-\tfrac{1}{2}\vartheta\phi_{2}\\
\phi_{4}^{\prime}= & \thinspace\phi_{4}+\tfrac{1}{2}\vartheta\phi_{5}\\
\phi_{5}^{\prime}= & \thinspace\phi_{5}-\tfrac{1}{2}\vartheta\phi_{4}\\
\phi_{6}^{\prime}= & \thinspace\phi_{6}-\tfrac{1}{2}\vartheta\phi_{7}\\
\phi_{7}^{\prime}= & \thinspace\phi_{7}+\tfrac{1}{2}\vartheta\phi_{6}
\end{aligned}
\right.~,\qquad\left\{ \begin{aligned}\psi_{0}^{\prime}= & \thinspace\psi_{0}-\vartheta\psi_{1}\\
\psi_{1}^{\prime}= & \thinspace\psi_{1}+\vartheta\psi_{0}\\
\psi_{2}^{\prime}= & \thinspace\psi_{2}\\
\psi_{3}^{\prime}= & \thinspace\psi_{3}\\
\psi_{4}^{\prime}= & \thinspace\psi_{4}\\
\psi_{5}^{\prime}= & \thinspace\psi_{5}\\
\psi_{6}^{\prime}= & \thinspace\psi_{6}\\
\psi_{7}^{\prime}= & \thinspace\psi_{7}
\end{aligned}
\right.~.
\end{equation}
What's peculiar here is the ways in which vector $x$ and spinors
$\phi$ and $\psi$ transform have interchanged between the three,
namely $x$ and $\phi$ appear to behave like a spinors now and $\psi$
looks like a vector, since full rotation in $\psi$ gives half a rotation
in $x$ and $\phi$. This is the property of the eight dimensional
space, which was named as \textit{triality}, similar to the \textit{duality}
for dual vector spaces.

For the completeness let us also write out boost-like non-compact
transformations, which are only realized in anisotropic spaces, for
example the transformations generated by $L_{04}\left(\vartheta\right)$,
\begin{equation}
\left\{ \begin{aligned}x_{0}^{\prime}= & \thinspace x_{0}+\vartheta x_{4}\\
x_{1}^{\prime}= & \thinspace x_{1}\\
x_{2}^{\prime}= & \thinspace x_{2}\\
x_{3}^{\prime}= & \thinspace x_{3}\\
x_{4}^{\prime}= & \thinspace x_{4}+\vartheta x_{0}\\
x_{5}^{\prime}= & \thinspace x_{5}\\
x_{6}^{\prime}= & \thinspace x_{6}\\
x_{7}^{\prime}= & \thinspace x_{7}
\end{aligned}
\right.~,\qquad\left\{ \begin{aligned}\phi_{0}^{\prime}= & \thinspace\phi_{0}-\tfrac{1}{2}\vartheta\phi_{4}\\
\phi_{1}^{\prime}= & \thinspace\phi_{1}-\tfrac{1}{2}\phi_{5}\vartheta\\
\phi_{2}^{\prime}= & \thinspace\phi_{2}-\tfrac{1}{2}\vartheta\phi_{6}\\
\phi_{3}^{\prime}= & \thinspace\phi_{3}-\tfrac{1}{2}\vartheta\phi_{7}\\
\phi_{4}^{\prime}= & \thinspace\phi_{4}-\tfrac{1}{2}\phi_{0}\vartheta\\
\phi_{5}^{\prime}= & \thinspace\phi_{5}-\tfrac{1}{2}\vartheta\phi_{1}\\
\phi_{6}^{\prime}= & \thinspace\phi_{6}-\tfrac{1}{2}\phi_{2}\vartheta\\
\phi_{7}^{\prime}= & \thinspace\phi_{7}-\tfrac{1}{2}\vartheta\phi_{3}
\end{aligned}
\right.~,\qquad\left\{ \begin{aligned}\psi_{0}^{\prime}= & \thinspace\psi_{0}-\tfrac{1}{2}\vartheta\psi_{4}\\
\psi_{1}^{\prime}= & \thinspace\psi_{1}+\tfrac{1}{2}\vartheta\psi_{5}\\
\psi_{2}^{\prime}= & \thinspace\psi_{2}+\tfrac{1}{2}\vartheta\psi_{6}\\
\psi_{3}^{\prime}= & \thinspace\psi_{3}+\tfrac{1}{2}\vartheta\psi_{7}\\
\psi_{4}^{\prime}= & \thinspace\psi_{4}-\tfrac{1}{2}\vartheta\psi_{0}\\
\psi_{5}^{\prime}= & \thinspace\psi_{5}+\tfrac{1}{2}\vartheta\psi_{1}\\
\psi_{6}^{\prime}= & \thinspace\psi_{6}+\tfrac{1}{2}\vartheta\psi_{2}\\
\psi_{7}^{\prime}= & \thinspace\psi_{7}+\tfrac{1}{2}\vartheta\psi_{3}
\end{aligned}
\right.~.
\end{equation}
We see that, similar to the compact case, the hyperbolic transformation
of one of the three objects (vector and two chiral spinors) in the
isotropic plane $e_{0}\wedge e_{4}$ generates spinorial transformations
of other two objects in corresponding four isotropic planes, $e_{0}\wedge e_{4}$,
$e_{1}\wedge e_{5}$, $e_{2}\wedge e_{6}$ and $e_{3}\wedge e_{6}$.
Again, it is possible to replicate transformations of $x$ in one
of the spinors which would trially swap their behavior.


\section{Split octonions}

\label{split} \setcounter{equation}{0}

It is known that spinors and vectors of (4+4)-space, considered in
Sec.~\ref{sec:vectors} and Sec.~\ref{sec:spinors}, can also be
represented using split octonions instead of matrices \cite{Nash-1,Nash-2,Nash-3,DeAndrade:2000jx}.
Split octonions $\mathbb{O}^{\prime}$ form non-associative algebra
with the property of alternativity. The algebra can be defined through
the algebraic relations: 
\begin{equation}
\begin{split}I^{2} & =1~,\qquad j_{n}I=J_{n}~,\qquad j_{m}j_{n}=-\delta_{mn}+\sum_{\ell}\epsilon_{\ell mn}j_{\ell}~,\qquad\qquad\left(\ell,m,n,=1,2,3\right)\\
J_{m}J_{n} & =\delta_{mn}+\sum_{\ell}\epsilon_{\ell mn}j_{\ell}~,\qquad J_{m}j_{n}=\delta_{mn}I-\sum_{\ell}\epsilon_{\ell mn}J_{\ell}~.
\end{split}
\label{eq:bassis-algebra}
\end{equation}
From the above relations and the alternativity property one can extract
the entire multiplication table for basis units 
\begin{equation}
\begin{array}{c||c|ccc|c|ccc}
 & 1 & j_{1} & j_{2} & j_{3} & I & J_{1} & J_{2} & J_{3}\\
\hline\hline 1 & 1 & j_{1} & j_{2} & j_{3} & I & J_{1} & J_{2} & J_{3}\\
\hline j_{1} & j_{1} & -1 & j_{3} & -j_{2} & J_{1} & -I & -J_{3} & J_{2}\\
j_{2} & j_{2} & -j_{3} & -1 & j_{1} & J_{2} & J_{3} & -I & -J_{1}\\
j_{3} & j_{3} & j_{2} & -j_{1} & -1 & J_{3} & -J_{2} & J_{1} & -I\\
\hline I & I & -J_{1} & -J_{2} & -J_{3} & 1 & -j_{1} & -j_{2} & -j_{3}\\
\hline J_{1} & J_{1} & I & -J_{3} & J_{2} & j_{1} & 1 & j_{3} & -j_{2}\\
J_{2} & J_{2} & J_{3} & I & -J_{1} & j_{2} & -j_{3} & 1 & j_{1}\\
J_{3} & J_{3} & -J_{2} & J_{1} & I & j_{3} & j_{2} & -j_{1} & 1
\end{array}\label{table:splitoct_multiplication}
\end{equation}

Now we want to show that complete algebra of the seven hypercomplex
basis units of the split octonions follows from the Moufang and Malcev
relations written for only three vector-like split octonionic elements
$J_{n}$. It is known that the anti-commuting basis units of octonions
and split octonions, $xy=-yx$, are Moufang loops \cite{Paal}. The
algebra formed by them is not associative but instead is alternative,
i.e. the associator 
\begin{equation}
\mathcal{A}(x,y,z)=\frac{1}{2}\bigg((xy)z-x(yz)\bigg)
\end{equation}
is totally antisymmetric 
\begin{equation}
\mathcal{A}(x,y,z)=-\mathcal{A}(y,x,z)=-\mathcal{A}(x,z,y)~.
\end{equation}
Consequently, any two units $x$ and $y$ generate an associative
subalgebra and obey the following mild associative laws: 
\begin{equation}
(xy)y=xy^{2}~,\quad x(xy)=x^{2}y~,\quad(xy)x=x(yx)~.
\end{equation}
The octonionic basis units also satisfy the flexible Moufang identities:
\begin{equation}
(xy)(zx)=x(yz)x~,\quad(zyz)x=z(y(zx))~,\quad x(yzy)=((xy)z)y~.
\end{equation}
In the algebra we have the following relationship between the associator,
\begin{equation}
\mathcal{A}(x,y,z)=\frac{1}{3}\bigg([x,[y,z]]+[y,[z,x]]+[z,[x,y]]\bigg)~,
\end{equation}
and the commutator 
\begin{equation}
[x,y]=\frac{1}{2}(xy-yx)~.
\end{equation}
Since the hypercomplex octonionic basis units anti-commute, their
commutator can always be replaced by the simple product, $[x,y]=xy$.

It is also known that basis units of octonions and split octonions
form the Malcev algebra (see, for example \cite{Carrion:2010it,Gunaydin:2016axc}).
Due to non-associativity, commutator algebra of split octonionic units
is non-Lie and instead of satisfying the Jacobi identity, they satisfy
the Malcev relation: 
\begin{equation}
(xy)(xz)=((xy)z)x+((yz)x)x+((zx)x)y~,
\end{equation}
or equivalently 
\begin{equation}
\mathcal{J}(x,y,(xz))=\mathcal{J}(x,y,z)x~,
\end{equation}
where 
\begin{equation}
\mathcal{J}(x,y,z)=\frac{1}{3}\bigg((xy)z+(yz)x+(zx)y\bigg)
\end{equation}
is so-called Jacobiator of $x$, $y$ and $z$. Indeed, using anti-commutativity
of elements, we find: 
\begin{equation}
\begin{split}3\,\mathcal{J}(x,y,(xz))=(xy)(xz)+(y(xz))x+((xz)x)y=\\
=((xy)z)x+((yz)x)x+((zx)x)y+(y(xz))x+((xz)x)y=\\
=((xy)z+(yz)x+(zx)y)x=3\,\mathcal{J}(x,y,z)x~.
\end{split}
\end{equation}

In Malcev's algebra two types of products are defined: bilinear $xy=-yx$
and trilinear $\mathcal{J}(x,y,z)$, which can be expressed using
bilinear products as: 
\begin{equation}
\mathcal{J}(x,y,z)=\frac{1}{3}\bigg(x(yz)+y(zx)+z(xy)\bigg)=-\mathcal{J}(y,x,z)=\mathcal{J}(x,z,y)~.
\end{equation}
We also have identities containing four and five elements of the algebra:
\begin{equation}
\begin{split}\mathcal{J}(xy,z,w) & +\mathcal{J}(yz,x,w)+\mathcal{J}(zx,y,w)=0~,\\
\mathcal{J}(x,y,zw) & =\mathcal{J}(x,y,z)w+z\mathcal{J}(x,y,w)~,\\
\mathcal{J}(x,y,\mathcal{J}(z,u,v)) & =\mathcal{J}(\mathcal{J}(x,y,z),u,v)+\mathcal{J}(z,\mathcal{J}(x,y,u),v)+\\
 & +\mathcal{J}(z,u,\mathcal{J}(x,y,v))~.
\end{split}
\end{equation}
So, one can generate a complete basis of the split octonions by the
multiplication and distribution laws of only three vector-like elements
$J_{n}$. Indeed, we can define pseudo-vector like basis units of
the split octonions $j_{n}$ by the commutators (or simple binary
products) of $J_{n}$, 
\begin{equation}
j_{n}=\frac{1}{2}\sum_{m}\sum_{k}\varepsilon_{nmk}J_{m}J_{k}~,\label{jI}
\end{equation}
where $\varepsilon_{nmk}$ is the totally antisymmetric unit tensor.
Also using Moufang identities for $J_{1}$, $J_{2}$ and $J_{3}$
we can identify the seventh basis unit $I$ with the Jacobiator, 
\begin{equation}
J_{1}j_{1}=J_{2}j_{2}=J_{3}j_{3}=-\mathcal{J}(J_{1},J_{2},J_{3})=I~.
\end{equation}
As a result, from the Moufang and Malcev relations we can recover
the complete algebra of all seven hypercomplex split octonionic basis
units (\ref{eq:bassis-algebra}). The non-vanishing associators of
these basis units are: 
\begin{equation}
\begin{split}\mathcal{A}(j_{n},J_{m},J_{k})=\delta_{nm}j_{k}-\delta_{nk}j_{m}~,\qquad\mathcal{A}(j_{n},j_{m},J_{k})=-\varepsilon_{nmk}I-\delta_{nk}J_{m}+\delta_{mk}J_{n}~,\\
\mathcal{A}(j_{n},j_{m},I)=\sum_{k}\varepsilon_{nmk}J_{k}~,\qquad\mathcal{A}(j_{n},J_{m},I)=-\sum_{k}\varepsilon_{nmk}j_{k}~,\\
\mathcal{A}(J_{n},J_{m},J_{k})=-\varepsilon_{nmk}I~,\qquad\mathcal{A}(J_{n},J_{m},I)=\sum_{k}\varepsilon_{nmk}J_{k}~.
\end{split}
\end{equation}

General split octonion $x\in\mathbb{O}^{\prime}$ over the field of
real numbers and its conjugate are 
\begin{equation}
\begin{split}x & =x_{0}+Ix_{4}+\sum_{n}\left(j_{n}x_{n}+J_{n}x_{4+n}\right)~,\\
\overline{x} & =x_{0}-Ix_{4}-\sum_{n}\left(j_{n}x_{n}+J_{n}x_{4+n}\right)~,
\end{split}
\end{equation}
where $n=1,2,3$ and $x_{0},x_{1},\ldots,x_{7}\in\mathbb{R}$. Quadratic
form $\mathcal{Q}:\mathbb{O}^{\prime}\rightarrow\mathbb{R}$ is defined
as multiplication of $x\in\mathbb{O}^{\prime}$ with its conjugate
\begin{equation}
\mathcal{Q}\left(x\right)=\overline{x}x~.
\end{equation}
The quadratic form cannot be used to construct a norm since it's not
positive definite and also evaluates to zero for nonzero split octonions.
Symmetric and non-degenerate bilinear form $\left\langle \cdot,\cdot\right\rangle :\mathbb{O}^{\prime}\times\mathbb{O}^{\prime}\rightarrow\mathbb{R}$
is defined in terms of the quadratic form as 
\begin{equation}
\left\langle x,y\right\rangle =\frac{1}{2}\mathcal{Q}\left(x+y\right)-\frac{1}{2}\mathcal{Q}\left(x\right)-\frac{1}{2}\mathcal{Q}\left(y\right).
\end{equation}
Explicitly it is 
\begin{equation}
\left\langle x,y\right\rangle =\frac{1}{2}\left(\overline{x}y+\overline{y}x\right)=\sum_{n=0}^{3}\left(x_{n}y_{n}-x_{4+n}y_{4+n}\right)~.\label{eq:splitoct_bilinear_form}
\end{equation}

At the end of this section we define split octonionic gradients: 
\begin{equation}
\begin{split}\partial & =\frac{1}{2}\left(\partial_{0}+I\partial_{4}\right)+\frac{1}{2}\sum_{n=0}^{3}\left(j_{n}\partial_{n}+J_{n}\partial_{4+n}\right)~,\\
\overline{\partial} & =\frac{1}{2}\left(\partial_{0}-I\partial_{4}\right)-\frac{1}{2}\sum_{n=0}^{3}\left(j_{n}\partial_{n}+J_{n}\partial_{4+n}\right)~,
\end{split}
\label{eq:splitoct_derivatives}
\end{equation}
where $\partial_{n}$ is a partial differentiation operator with respect
to $x_{n}$. They are defined in such a way that they mimic properties
of regular derivative for $\mathbb{R}\rightarrow\mathbb{R}$ functions
and Wirtinger derivatives for $\mathbb{C}\rightarrow\mathbb{C}$ functions,
namely 
\begin{equation}
\begin{split}\partial x=\overline{\partial}\overline{x}=1~,\\
\overline{\partial}x=\partial\overline{x}=0~.
\end{split}
\end{equation}
But but these properties do not extend to higher order terms in $x$
and $\overline{x}$ \cite{KauhanenOrelma_OctonionicAnalysis}, since
they already fail for quaternionic derivatives \cite{DeLeoRotelli_QuaternoinicAnalysis}.


\section{Split octonions and triality}

\label{sec:Trilinear} \setcounter{equation}{0}

Now let us express the triality of (4+4)-space in terms of split octonions.
We can write split octonionic representation of the (4+4)-space vector
and chiral spinors, (\ref{eq:cl44vector}) and (\ref{eq:cl44chiral_spinors}),
as 
\begin{equation}
\begin{split}\phi & =\phi_{0}+\phi_{1}j_{1}+\phi_{2}j_{2}+\phi_{3}j_{3}+\phi_{4}I+\phi_{5}J_{1}+\phi_{6}J_{2}+\phi_{7}J_{3}~,\\
x & =x_{0}+x_{1}j_{1}+x_{2}j_{2}+x_{3}j_{3}+x_{4}I+x_{5}J_{1}+x_{6}J_{2}+x_{7}J_{3}~,\\
\psi & =\psi_{0}+\psi_{1}j_{1}+\psi_{2}j_{2}+\psi_{3}j_{3}+\psi_{4}I+\psi_{5}J_{1}+\psi_{6}J_{2}+\psi_{7}J_{3}~.
\end{split}
\label{eq:X_phi_psi}
\end{equation}
Note that unlike the Clifford algebraic representation of spinors
and vectors (considered in Sec.~\ref{sec:vectors} and Sec.~\ref{sec:spinors}),
where they are represented by different type of objects, here they
are a same type of object. Furthermore, the invariants constructed
by the split octonionic vector and spinors (\ref{eq:X_phi_psi}),
can also be written identically to each other, namely 
\begin{equation}
\mathcal{Q}\left(\phi\right)=\overline{\phi}\phi~,\qquad\mathcal{Q}\left(x\right)=\overline{x}x~,\qquad\mathcal{Q}\left(\psi\right)=\overline{\psi}\psi~.
\end{equation}
These expressions respect the fact that they evaluate to same quadratic
forms (\ref{eq:cl44quadratic_form}) and (\ref{eq:cl44spinors_quadratic_forms})
and are interchangeable in the trilinear form (\ref{F_matrixrep})
as we have seen above.

Trilinear form (\ref{F_matrixrep}) represented with split octonions
can be written using bilinear form (\ref{eq:splitoct_bilinear_form})
as \cite{Gurchumelia:2021yky}, 
\begin{equation}
\mathcal{F}\left(\phi,x,\psi\right)=\left\langle \overline{\phi},x\psi\right\rangle \thinspace.\label{Trilinear}
\end{equation}

Split octonionic vectorial and spinorial representations of $SO(4,4)$
can be constructed similarly to octonionic representations of $SO(8)$
as described in \cite{Dray}. Let us denote two distinct basis units
of split octonions by $u$ and $v$ and define an object: 
\begin{equation}
T_{uv}=\begin{cases}
u\cos\left(\frac{\vartheta}{2}\right)+v\sin\left(\frac{\vartheta}{2}\right)~, & u\overline{u}=v\overline{v}\\
u\cosh\left(\frac{\vartheta}{2}\right)+v\sinh\left(\frac{\vartheta}{2}\right)~, & u\overline{u}=-v\overline{v}
\end{cases}~.
\end{equation}
Then any group transformation can be constructed by compositions of
transformations 
\begin{equation}
x^{\prime}=T_{uv}\left(uxu\right)T_{uv}\label{eq:splitoctSO44}
\end{equation}
for $SO\left(4,4\right)$ and 
\begin{equation}
\phi{}^{\prime}=\left(\phi u\right)\overline{T}_{uv}~,\label{eq:splitoct_Spin44_phi}
\end{equation}
\begin{equation}
\psi{}^{\prime}=\overline{T}_{uv}\left(u\psi\right)~,\label{eq:splitoct_Spin44_psi}
\end{equation}
for $Spin\left(4,4\right)$. For example, the transformation $L_{01}(\vartheta)$,
whose infinitesimal version is written out in (\ref{eq:L_01}), is
achieved by taking $u=1$ and $v=j_{1}$ in (\ref{eq:splitoctSO44}),
(\ref{eq:splitoct_Spin44_phi}) and (\ref{eq:splitoct_Spin44_psi}),
since $1$ and $j_{1}$ are the first two units of split octonions.
Ordering of the rest of the basis units is same as in the multiplication
table (\ref{table:splitoct_multiplication}).


\section{Split octonionic field theories}

\label{sec:Field_theory} \setcounter{equation}{0}

Using the expression of trilinear form (\ref{Trilinear}) we can construct
a Lagrangian by replacing $x\in\mathbb{O}^{\prime}$ with the derivative
defined in (\ref{eq:splitoct_derivatives}) that act to the right,
\begin{equation}
\mathcal{L}=\left\langle \overline{\phi},\overrightarrow{\partial}\psi\right\rangle ~.
\end{equation}
By stationarizing the action integral 
\begin{equation}
S=\int d^{8}x\mathcal{L}~,
\end{equation}
we get right-analyticity and left-analyticity conditions on $\phi$
and $\psi$ 
\begin{equation}
\left\{ \begin{aligned}\phi\overleftarrow{\partial}= & 0~,\\
\overrightarrow{\partial}\psi= & 0~.
\end{aligned}
\right.\label{eq:leftnright_splitoct_analyticity}
\end{equation}
These equations represent the generalized Cauchy-Riemann and Cauchy-Riemann-Fueter
\cite{Sudbery_CauchyRiemannFueterEq} conditions for split octonions.
Split octonionic conjugation of the first equation in (\ref{eq:leftnright_splitoct_analyticity})
turns the system into 
\begin{equation}
\left\{ \begin{aligned}\overrightarrow{\overline{\partial}}\overline{\phi}= & 0~,\\
\overrightarrow{\partial}\psi= & 0~.
\end{aligned}
\right.
\end{equation}

Adding quadratic terms in $\phi$ and $\psi$ to the Lagrangian 
\begin{equation}
\mathcal{L}=\left\langle \overline{\phi},\overrightarrow{\partial}\psi\right\rangle +\frac{1}{2}\lambda_{1}\left\langle \phi,\phi\right\rangle +\frac{1}{2}\lambda_{2}\left\langle \psi,\psi\right\rangle \label{eq:quadratic_lagrangian}
\end{equation}
results in mixing of the these split octonionic fields at the equation
level, 
\begin{equation}
\left\{ \begin{aligned}\overrightarrow{\overline{\partial}}\overline{\phi}= & \lambda_{2}\psi~,\\
\overrightarrow{\partial}\psi= & -\lambda_{1}\overline{\phi}~.
\end{aligned}
\right.
\end{equation}
If we take $\lambda_{2}=0$ and use the property of alternativity,
then equations reduce to eight independent Klein-Gordon like equations
for (4+4)-space 
\begin{equation}
\left\langle \overrightarrow{\partial},\overrightarrow{\partial}\right\rangle \psi=0\thinspace.
\end{equation}


\section{Dirac and Maxwell equations}

\label{sec:Dirac+Maxwell} \setcounter{equation}{0}

We define a new gradient operator $D$ in terms of $\partial$ as
\begin{equation}
D=I\partial I~,
\end{equation}
which has the opposite sign for imaginary parts $j_{n}$ and $J_{n}$
\begin{equation}
D=\frac{1}{2}\left(\partial_{0}+I\partial_{4}\right)-\frac{1}{2}\sum_{n}\left(j_{n}\partial_{n}+J_{n}\partial_{4+n}\right)~.
\end{equation}
We also consider fields (the case of split quaternions see in \cite{Gogberashvili:2022opk,Gogberashvili:2021pel}):
\begin{equation}
A=\mathcal{C}_{0}+j_{1}\mathcal{A}_{1}+j_{2}\mathcal{A}_{2}+j_{3}\mathcal{A}_{3}+I\mathcal{A}_{0}+J_{1}\mathcal{C}_{1}+J_{2}\mathcal{C}_{2}+J_{3}\mathcal{C}_{3}~,
\end{equation}
\begin{equation}
F=\overrightarrow{D}A~.\label{eq:splitoct_fieldF}
\end{equation}
If we take the quadratic Lagrangian defined above (\ref{eq:quadratic_lagrangian})
with parameters $\lambda_{2}=0$ and $\lambda_{1}=-1$, set $\phi=\overline{F}$
and $\psi=A$ and use $D$ instead of $\partial$ we get 
\begin{equation}
\mathcal{L}=\left\langle F,\overrightarrow{D}A\right\rangle -\frac{1}{2}\left\langle \overline{F},\overline{F}\right\rangle \thinspace.
\end{equation}
Using the definition of $F$ in terms of $A$ (\ref{eq:splitoct_fieldF})
and the fact that $\left\langle \overline{F},\overline{F}\right\rangle =\left\langle F,F\right\rangle $,
the Lagrangian simplifies to 
\begin{equation}
\mathcal{L}=\frac{1}{4}\left\langle F,F\right\rangle ~.
\end{equation}
Equation of motion for this Lagrangian is 
\begin{equation}
\left\langle \overrightarrow{D},\overrightarrow{D}\right\rangle A=0~.\label{eq:SplitOct_44DyonicMaxwell}
\end{equation}
In the limit when $D\rightarrow\mathscr{D}=\frac{1}{2}\left(-j_{1}\partial_{x}-j_{2}\partial_{y}-j_{3}\partial_{z}+I\partial_{t}\right)$
the equation (\ref{eq:SplitOct_44DyonicMaxwell}) reduces to free
dyonic Maxwell equations in Minkowski space 
\begin{equation}
\left\langle \overrightarrow{\mathscr{D}},\overrightarrow{\mathscr{D}}\right\rangle A=0\thinspace,
\end{equation}
where $\mathcal{A}_{n}$ and $\mathcal{C}_{n}$ for $n=0,1,2,3$ are
electromagnetic and dyonic 4-potentials, split octonionic form of
which was first introduced in \cite{ChanyalB2014,Chanyal:2010sz}.

For the following Lagrangian with mass parameter $m$
\begin{equation}
\mathcal{L}=\left\langle \overline{\phi},\overrightarrow{D}\psi\right\rangle -\frac{1}{2}m\left\langle \overline{\phi},J_{3}\psi\right\rangle =\left\langle \overline{\phi},\left(\overrightarrow{D}-\frac{1}{2}mJ_{3}\right)\psi\right\rangle ~,
\end{equation}
two independent equations of motion result after stationarizing the
corresponding action 
\begin{equation}
\left\{ \begin{aligned}\left(\overrightarrow{\overline{D}}-\frac{1}{2}J_{3}m\right)\overline{\phi}= & 0~,\\
\left(\overrightarrow{D}-\frac{1}{2}J_{3}m\right)\psi= & 0~,
\end{aligned}
\right.\label{eq:diraclike_eom}
\end{equation}
second of which reduces to Dirac equation in the limit $D\rightarrow\mathscr{D}$.

Lagrangian for a single field $\psi$ obtained by setting $\phi=\overline{\psi}J_{3}$
and taking the limit $D\rightarrow\mathscr{D}$

\begin{equation}
\mathcal{L}=\left\langle -J_{3}\psi,\left(\overrightarrow{\mathscr{D}}-\frac{1}{2}J_{3}m\right)\psi\right\rangle 
\end{equation}
is equivalent to Dirac Lagrangian and consequently resulting equation
of motion
\begin{equation}
\overrightarrow{\mathscr{D}}\psi=\frac{1}{2}J_{3}m\psi\thinspace,
\end{equation}
is equivalent to the regular Dirac equation.


\section{Summary and concluding remarks}

\label{sec:conclusions}

In this paper the equivalence of 8-dimensional spinors and vectors
is discussed for (4+4)-space within the context of the algebra of
the split octonions. It is shown that the complete algebra of hypercomplex
split octonionic basis units can be recovered from the Moufang and
Malcev relations for the three vector-like elements. The trilinear
form, together with $SO(4,4)$ and $Spin(4,4)$ group transformations,
under which it is invariant, is represented using split octonions.
It is shown that, unlike matrix cases, this representation respects
the triality symmetry. Subsequently Lagrangians on split octonionic
spinorial and vectorial fields were constructed using the group invariant
quadratic and trilinear forms. Split octonionic analyticity conditions
were obtained by stationarizing action corresponding to the simplest
trilinear Lagrangian. It is shown that similar Lagrangians correspond
to the system that reduces to the free Dirac and dyonic Maxwell equations
when extra four spacetime dimensions are removed. It is worth noting
that the trilinear relation is exactly of the form used in supersymmetric
theories (see, for example \cite{Schray:1994up,Anastasiou:2013cya}),
so it is only natural that the overall symmetry of such models is
given by triality algebras.


\section*{Declarations:}

\subsection*{Ethical Approval}

Not applicable.

\subsection*{Authors' contributions}

Both M.G. and A.G. contributed to the final version of the manuscript.
M.G. supervised the project.

\subsection*{Funding}

This work was partially supported by the joint grant of Volkswagen
Foundation and SRNSF (Ref. 93 562 \& \#04/48).

\subsection*{Availability of data and materials}

The source code for symbolic computations used during the current study along with calculations in
Jupyter notebooks are publicly available at: https://github.com/EQUINOX24/SplitOct


\end{document}